\documentclass[pra,reprint,superscriptaddress,amsmath,amssymb,aps]{revtex4-1}

\usepackage{hyperref}
\usepackage{graphicx}

\usepackage{relsize}
\usepackage{color}

\usepackage[normalem]{ulem} 	
\definecolor{purple}{rgb}{0.63,0,1}
\newcommand{\as}[1]{{{#1}}}

\definecolor{MyDarkGreen}{rgb}{0,0.45,0.08}
\newcommand{\mc}[1]{{#1}}

\definecolor{orange}{RGB}{222,93,18}

\definecolor{blue}{RGB}{26,0,255}

\usepackage[notext]{stix}
\usepackage{times}

\hypersetup{%
    unicode=true,
    colorlinks=true,
    linkcolor=blue,
    citecolor=blue,
    urlcolor=blue,
    pdftitle={Many-body correlations in one-dimensional optical lattices with alkaline-earth(-like) atoms},
    pdfauthor={V. Bilokon, E. Bilokon, M. C. Bañuls, A. Cichy, A. Sotnikov},
    pdfproducer={},
    pdfcreator={}}

\newcommand{\subfigref}[2]{\hyperref[fig:#1]{\ref*{fig:#1}(#2)}}

\begin{document}
\title{Many-body correlations in one-dimensional optical lattices with alkaline-earth(-like) atoms}

 \author{Valeriia Bilokon}
 \affiliation{V. N. Karazin Kharkiv National University, Svobody Sq. 4, 61022 Kharkiv, Ukraine}
 \affiliation{Institute of Spintronics and Quantum Information, Faculty of Physics, Adam Mickiewicz University in Poznań, Uniwersytetu Poznańskiego 2, 61-614 Poznań, Poland}
 \author{Elvira Bilokon}
 \affiliation{V. N. Karazin Kharkiv National University, Svobody Sq. 4, 61022 Kharkiv, Ukraine}
 \affiliation{Institute of Spintronics and Quantum Information, Faculty of Physics, Adam Mickiewicz University in Poznań, Uniwersytetu Poznańskiego 2, 61-614 Poznań, Poland}
 \author{Mari Carmen Bañuls}
 \affiliation{Max-Planck-Institut f{\"u}r Quantenoptik, Hans-Kopfermann-Stra{\ss}e 1, 85748 Garching, Germany}
 \affiliation{Munich Centre for Quantum Science and Technology (MCQST), Schellingstrasse 4, Munich 80799, Germany}
 \author{Agnieszka Cichy}
 \affiliation{Institute of Spintronics and Quantum Information, Faculty of Physics, Adam Mickiewicz University in Poznań, Uniwersytetu Poznańskiego 2, 61-614 Poznań, Poland}
 \affiliation{Institut f\"ur Physik, Johannes Gutenberg-Universit\"at Mainz, Staudingerweg 9, 55099 Mainz, Germany}
 \author{Andrii Sotnikov}
 \email{a\_sotnikov@kipt.kharkov.ua}
 \affiliation{V. N. Karazin Kharkiv National University, Svobody Sq. 4, 61022 Kharkiv, Ukraine}
 \affiliation{Kharkiv Institute of Physics and Technology, Akademichna 1, 61108 Kharkiv, Ukraine}
 
\date{\today}

\begin{abstract}
We explore the rich nature of correlations in the ground state of ultracold atoms trapped in state-dependent optical lattices.
In particular, we consider interacting fermionic ytterbium or strontium atoms, realizing a two-orbital Hubbard model with two spin components.
We analyze the model in one-dimensional setting with the experimentally relevant hierarchy of tunneling and interaction amplitudes by means of exact diagonalization and matrix product states approaches, and study the correlation functions in density, spin, and orbital sectors as functions of variable densities of atoms in the ground and metastable excited states. 
We show that in certain ranges of densities these atomic systems demonstrate strong density-wave, ferro- and antiferromagnetic, as well as antiferroorbital correlations.
\end{abstract}

\maketitle

\section{Introduction}

By means of near-resonant laser fields,  advances in cooling, trapping, and loading neutral atoms in optical lattices have allowed a detailed study of unique properties of quantum many-body systems.
A major appeal of these studies is the possibility of realizing strongly-correlated phases. They are interesting not only on their own, but also from the viewpoint of using atoms in optical lattices as universal quantum simulators of electrons in crystalline materials~\cite{Gross17}. 
In turn, a key feature of electrons in strongly-correlated solid-state compounds, e.g., in the transition-metal oxides, is the fermions possession of spin and orbital degrees of freedom, which are equally vital for many emerging phenomena.
Therefore, realization and control over many-body systems with the capabilities for all relevant internal degrees of freedom of ``elementary'' particles become highly important.

Recently, a two-orbital Fermi-Hubbard system has been realized with ultracold alkaline-earth(-like) atoms (AEAs) in a state-dependent optical lattice (SDL)~\cite{Riegger18,Heinz2020,Ono21}.
This model has been a subject of many theoretical studies due to additional unique phenomena peculiar to multiorbital lattice systems.
In certain limits it reveals, in particular, the physics of the orbital-selective Mott transition \cite{PhysRevB.87.205135}, the Kugel-Khomskii model originally proposed for transition-metal oxides \cite{Sov.Phys.Usp.25.231}, the Kondo lattice model \cite{FossFeig10} studied in the context of manganese oxide perovskites and heavy fermion materials, and SU$(N)$-symmetric magnetic systems \cite{Gorshkov2010NP,Rep.Prog.Phys.77.124401}.
Now, important questions appear about the optimal regimes for the realization of particular strongly-correlated phenomena within novel cold-atom systems. 
In this paper, we address the mentioned questions by performing theoretical analysis of the two-orbital Fermi-Hubbard model with experimentally-relevant parameters corresponding to particular AEAs and quasi-one-dimensional geometry of SDL.
Compared to previous theoretical studies of the one-dimensional Fermi-Hubbard model with two orbital and two spin flavors (see, e.g., Refs.~\cite{Miyashita2009, Plekhanov2011, Nonne2010, Bois2015, Capponi2016},), here we systematically explore ground-state properties of AEA systems in a wide range of lattice fillings and experimentally relevant interaction and hopping amplitudes.

\section{System, Model, and Methods}\label{sec:system-model-method}

\subsection{Fermionic isotopes of Yb and Sr atoms in state-dependent optical lattices}\label{sec:system}
Our research is motivated by recent developments in  experiments with ultracold gases of alkaline-earth(-like) atoms. 
These atomic systems, in particular the fermionic isotopes of strontium and ytterbium ($^{87}$Sr, $^{171}$Yb, and  $^{173}$Yb) set previously unexplored perspectives for the investigation of new states of matter \cite{Gorshkov2010NP,FossFeig10,Rep.Prog.Phys.77.124401}. 
\as{In this respect, they offer advantages over the more traditionally used alkali-metal atoms by possessing, in particular,} two key properties: (i) the existence of a long-lived metastable $^3$P$_0$ electronic state (denoted below as $e$) coupled to the $^1$S$_0$ ground state (denoted below as $g$) through an ultranarrow optical transition and (ii) the vanishing electronic angular momentum ($J=0$) in both of these states. The metastable state offers an additional degree of freedom, since its interaction properties -- both with light and with other states -- differ strongly from the ground state.   This allows experimental realization of the two-band Hubbard model. 

The study focuses on ultracold gases of strontium or ytterbium atoms being prepared in two different orbital states $|g\rangle$ and  $|e\rangle$, and two different nuclear (pseudo-)spin states 
$|{\uparrow}\rangle$ and $|{\downarrow}\rangle$.
Thanks to successful experiments with measurements of interactions for all three fermionic isotopes: $^{87}$Sr, $^{171}$Yb and $^{173}$Yb, where the $s$-wave scattering amplitudes for intra- and inter-orbital interactions were relatively well determined \cite{Zhang14S, Goban2018, Kitagawa08, Bettermann2020, Ono21, Scazza2014NP, Hoefer15}, we can summarize these in Table~\ref{tbl:s_lengths}.
\as{Note that the given scattering lengths have different relative magnitudes (in particular, one can realize both $a_{eg}^{+}>a_{eg}^{-}$ and $a_{eg}^{+}<a_{eg}^{-}$, moderate or vanishing $a_{gg}$, etc.).}
This means that every atomic system can be unique and important for an enhancement or suppression of specific many-body correlations in certain regimes.
\begin{table}
\begin{center}
\begin{tabular}{|c|cccc|c|}
   \hline
                & $a_{gg}$  & $a_{ee}$  & $a_{eg}^+$ & $a_{eg}^-$ & Refs. \\[1mm]
   \hline
   $^{173}$Yb   & 199.4     & 306.2     & 1878 & 220 & \cite{Kitagawa08, Scazza2014NP, Hoefer15}
   \\
   $^{171}$Yb   & $\approx0$ & 104      & 240 & 389 & \cite{Kitagawa08, Bettermann2020, Ono21}
   \\
   $^{87}$Sr    & 96.2     & 176.0     & 169 & 68 & \cite{Zhang14S,Goban2018}
   \\
   \hline
\end{tabular}
\end{center}
\caption{Intra- and inter-orbital $s$-wave scattering lengths of fermionic AEL atoms in units of the Bohr radius~$a_0$ with the representative references (for more detailed information on measurements, see also the references therein).}\label{tbl:s_lengths}
\end{table}

In general, the near-resonant laser field with a certain wavelength~$\lambda$ creating the optical lattice interacts differently with atoms in the states $|g\rangle$ and  $|e\rangle$, thus the lattice can be viewed as state-dependent. For the three considered isotopes, one can determine a particular ``magic'' wavelength~$\lambda_{\rm m}$ at which atoms in two orbital states have equal polarizabilities, i.e., the lattice depth becomes equal for both orbital components. Below, we use both the magic-wavelength and SDL options. In particular, we
set that the state-dependent lattice with a mo\-de\-ra\-te amplitude is created along one spatial direction, while a stronger confinement via the magic-wavelength optical lattice is acting in transversal directions. In this respect, the system can be viewed as effectively quasi-one-dimensional.
For de\-fi\-ni\-te\-ness, we assume that SDL is created along the $x$ direction and has a moderate amplitude $\as{V_{x}^{(g)}}=5E_r$, where $E_r=\hslash^2 k^2/2m$ is the recoil energy of an atom with the mass $m$ and $\hslash$ is Planck's constant and $k=2\pi/\lambda$.
To have a certain correspondence with previous theoretical studies~\cite{Sotnikov2020PRR} and for convenience of the analysis, we choose the polarizability ratio to be equal for all atoms, $p=2.1$\as{, also meaning that $V_{x}^{(e)}=10.5E_r$} \footnote{According to the additional analysis, the main results remain qualitatively similar at different values of the polarizability ratio, $p=1.2$ and $p=3.3$, in particular.}. In particular, for ytterbium isotopes this results in $\lambda_{\rm SDL}\approx690$~nm \cite{Sotnikov2020PRR}, while for strontium atoms this yields $\lambda_{\rm SDL}\approx739$~nm \cite{Safronova2015}.
The state-independent (``magic-wavelength'') confinement is realized by taking $V_y=V_z=18E_r$ (with $\lambda_{\rm m}\approx759$~nm \cite{Riegger18} and $\lambda_{\rm m}\approx813$~nm \cite{Takamoto2005} for Yb and Sr isotopes, respectively).

Below, we also focus on homogeneous (but finite-size) systems neglecting all effects originating from the trapping potential. These can be naturally included in the theoretical formalism, but the analysis of the effects related to additional inhomogeneities goes beyond the scope of the current study.

\subsection{Two-orbital Hubbard model and coupling amplitudes}
Within the tight-binding approximation, the system can be described by the two-orbital Hubbard model \as{\cite{FossFeig10,Gorshkov2010NP}}:
\begin{eqnarray}\label{2bhm}
    \mathcal{H} &=&  \sum_{i, \gamma, \sigma}t_{\gamma}
     (c_{i\gamma \sigma}^{\dag} c_{i+1\gamma \sigma}+{\rm H.c.})
     -\sum_{i,\gamma}\mu_{\gamma}n_{i\gamma}+ \mathcal{H}_{\rm int}, 
\end{eqnarray}
 where
\begin{eqnarray}\label{eq:Hint}
     \mathcal{H}_{\rm int}&=&\sum_{i, \gamma}U_{\gamma\gamma} \sum_{\sigma < \sigma'} n_{i \gamma \sigma}  n_{i \gamma \sigma'}
     + V \sum_{i,\sigma<\sigma ', \gamma<\gamma '} n_{i \gamma \sigma} n_{i \gamma' \sigma'}
     \\
     \nonumber
     &&
     + (V-V_{\rm ex}) \sum_{i,\sigma, \gamma < \gamma'} n_{i \gamma \sigma} n_{i \gamma' \sigma}
     \\
     \nonumber
     &&
     + V_{\rm ex} \sum_{i, \sigma<\sigma', \gamma < \gamma'} c_{i \gamma \sigma}^{\dag} c_{i \gamma' \sigma'}^{\dag} 
     c_{i \gamma \sigma'} c_{i \gamma' \sigma}. 
\end{eqnarray}
The indices $\gamma, \gamma'=\{g,e\}$ and $\sigma, \sigma'=\{\uparrow, \downarrow\}$ denote the orbital states and the nuclear Zeeman spin states, respectively. 
The operator $c_{i\gamma \sigma}^{\dag}$ ($c_{i\gamma \sigma}$) creates (annihilates) an atom in the internal state $|\gamma \sigma \rangle$ at the site $i=1,\dots, L$, where $L$ is the size of the chain. 
The local density operator of atoms in the orbital state $\gamma$ is $n_{i\gamma}=\sum_{\sigma} n_{i\gamma \sigma}$ and $n_{i\gamma \sigma}=c_{i\gamma \sigma}^{\dag} c_{i\gamma \sigma}$. For a particular orbital state $\gamma$, $t_{\gamma}$ is the hopping amplitude and $\mu_{\gamma}$ is the chemical potential.
\as{We should note that from the point of view of quantitative comparison with the experimental realizations, the model~\eqref{2bhm} can be viewed as a significant simplification. In particular, for the chosen lattice depth $V_{x}^{(g)}=5E_r$, other subleading terms as the density-assisted hopping \cite{Luhmann2012}, next-nearest neighbor hopping \cite{Blo2008RMP}, higher-band contributions, as well as effects originating from the degrees of freedom in transversal directions ($V_{y,z}=18E_r$) can alter observables on the quantitative scale. 
To confirm that our central results are robust against such perturbations, we performed additional analysis with the renormalized hopping amplitudes $t_g$, different lattice depths, and different values of the polarizability ratio $p$. 
}

The local interaction amplitudes within the lowest-band approximation for both $g$ and $e$ orbital states can be estimated by
\begin{equation}\label{Hubb_par}
    U_{\gamma \gamma'}=g_{\gamma \gamma'} \int d^3 r w_{\gamma}^{2}({\bf r}) w_{\gamma'}^{2}({\bf r}),  
\end{equation}
with $w_\gamma({\bf r})$ being the Wannier function of an atom in the orbital state $\gamma$, 
and the coupling $g_{\gamma\gamma'}=4\pi \hbar^2 a_{\gamma\gamma'}/m$, where $a_{\gamma\gamma'}$ is the scattering length of two atoms in the states $\gamma$ and $\gamma'$ (see Table~\ref{tbl:s_lengths}). 
For inter-orbital scattering, two different scattering lengths $a_{eg}^{\pm}$ (and correspondingly two amplitudes $U_{eg}^{\pm}$ computed as in \eqref{Hubb_par}) appear, for the triplet ($+$) or singlet  ($-$) configuration of the pair of atoms. In terms of them, on-site direct and exchange interactions are obtained respectively as $V=(U_{eg}^{+}+U_{eg}^{-})/2$ and $V_{\rm ex}=(U_{eg}^{+}-U_{eg}^{-})/2$. 
Note that the inter-orbital exchange interaction $V_{\rm ex}$ can be separated into its density-density and spin-flip contributions, see the terms in the second and the third lines of Eq.~(\ref{eq:Hint}), respectively.

\begin{figure}
\begin{center}
    \includegraphics[width=\linewidth]{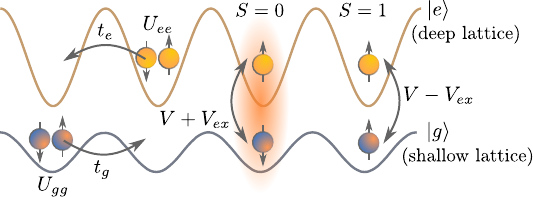}
    \caption{Sketch of hopping and interaction processes in the one-dimensional two-orbital Hubbard model with two different spin states ($\uparrow$ and $\downarrow$). 
    $U_{gg}$ and $U_{ee}$ are the intraorbital interactions between the atoms in the ground ($g$) and excited ($e$) states, respectively. 
    $V$ denotes the direct interaction between atoms in different orbital states, while $V_{\rm ex}$ represents indirect (exchange) interaction between them. 
    $t_g$ and $t_e$ correspond to the hopping amplitudes of atoms between the nearest-neighbor sites.} 
    \label{interactions}
\end{center}
\end{figure}
The one-dimensional system described by the Hamiltonian~(\ref{2bhm}) (see also Fig.~\ref{interactions}) can be experimentally realized with AEL atoms in the state-dependent optical lattices, as specified in Sec.~\ref{sec:system}. In the given form, the model is also closely related to solid-state realizations, since it contains all relevant interaction and hopping processes peculiar to electrons in two distinct orbital states in crystalline materials.

\subsection{Numerical approaches}\label{sec:methods}

Among theoretical approaches, the exact diagonalization (ED) provides a direct way to extract full information about eigenstates of the quantum many-body system with access to all the relevant physical observables, e.g., the local densities, double occupancies, spin-spin and orbital-orbital correlations, etc. 
However, the application of the ED approach is strongly limited by an exponential growth of the corresponding Hilbert space. 
Since in the model~\eqref{2bhm} there are four internal degrees of freedom of fermions per site, this sets a restriction to the system size $L\approx5$ available for a direct numerical analysis if no additional optimizations are applied.

Fortunately, the last decades brought a new generation of non-perturbative techniques
for numerical analysis of quantum many-body problems. 
Among them, tensor network (TN) methods~\cite{Verstraete2008,schollwock11,Orus2014annphys,Silvi2019tn,Okunishi2022,Banuls2023} provide efficient descriptions of quantum many-body strongly correlated states based on their entanglement properties.
The paradigmatic example of TN state is the 
matrix product state (MPS) ansatz~\cite{fannes92,vidal04,verstraete04}. 
MPS-based approaches capture the entanglement area law~\cite{Eisert2010} in one spatial dimension, 
and underlie the successful density matrix renormalization group (DMRG)~\cite{white,schollwock11} algorithm, 
state-of-the-art method for numerical quasi-exact solution of strongly-correlated problems in one dimension, 
which effectively minimizes the energy over the set of MPS.

Here, we optimize variationally an MPS ansatz to study the ground state of the two-orbital Fermi-Hubbard model for up to $L=40$ sites,
and analyze its many-body correlation functions. We compare and benchmark our results against ED results for smaller system sizes.
Note that the two-orbital Fermi-Hubbard model was also the subject of previous DMRG stu\-dies with the solid-state parametrization of the interaction amplitudes \cite{Miyashita2009,Plekhanov2011} and partially AEA-like parametrization at half-filling in Refs.~\cite{Nonne2010, Bois2015, Capponi2016}.
In contrast, here we employ a hierarchy of the interaction amplitudes dictated by the scattering lengths in the cold-atom realizations (see Table~\ref{tbl:s_lengths}) and perform systematic analysis in a wide range of atomic densities.

Whereas the details of the numerical method can be found in the literature~\cite{Verstraete2008, schollwock11}, let us briefly outline the main ingredients in the MPS approach for the system under study.
The MPS ansatz for the state of a quantum $N$-body system has the following form:
\begin{equation}\label{iterative}
    |\Psi\rangle = \sum_{i_1\ldots i_{L}=1}^{d} \textrm{tr}\left(A_1^{i_1}\ldots A_{L}^{i_{L}}\right)
    |i_1\ldots i_{L}\rangle,
\end{equation} 
where $|i_k\rangle$ are the single-site basis states
(with $d$ being the dimension of the single-site Hilbert space) and each $A_{k}^{i_k}$ is a $D\times D$ matrix, where $D$ is called the bond dimension \footnote{For open boundary conditions, as we consider here, the first and last matrices have dimensions $1\times D$ and $D\times 1$, respectively.}.

The MPS is a convenient ansatz for the ground state of local one-dimensional Hamiltonians. 
Although it is possible to use TN directly for fermionic systems~\cite{Kraus2010,Corboz2009fMERA,Corboz2010fpeps,Pineda2010}, for one-dimensional problems it is convenient to employ the Jordan--Wigner (JW) transformation~\cite{Jordan:1928wi} and map the original two-orbital Fermi--Hubbard model~(\ref{2bhm}) to the Hamiltonian of a spin chain.

In order to apply the JW transformation, we define a linear order for the fermionic modes $c_{i\gamma\sigma}$ according to the generalized index $m=4(i-1)+2(i_{\gamma}-1)+i_{\sigma}$ ($m=1,\ldots, 4L$), where $i_{\gamma,\sigma}=\{1,2\}$ number the internal orbital and spin fermionic modes on the site $i$. 
For later convenience, we also define the internal state linear index for each site $k=2(i_{\gamma}-1)+i_{\sigma}$, taking values $k=1,\ldots 4$.
The fermionic operators are thus mapped to strings of the spin-1/2 Pauli matrices as
\begin{equation}
    c^{\dagger}_m=\prod_{q=1}^{m-1}(-\sigma_{q}^{z})\cdot\sigma^+_m,
    \qquad
    c_m=\prod_{q=1}^{m-1}(-\sigma_{q}^{z})\cdot\sigma^-_m.
\end{equation}
Note that the density operator for a single fermionic mode~$\hat{n}_m$ can be written as
    $\hat{n}_m = c^{\dagger}_m c_m
    = \sigma^+_m\sigma^-_m
     = \pi^0_m,$
where $\pi^0$ is the projection operator
\begin{equation}
     \pi^0 = 
     \left(
     \begin{array}{cc}
      1 & 0
      \\
      0 & 0
     \end{array}
     \right)
    .
\end{equation}
Therefore, in terms of these matrices, we express the system Hamiltonian~(\ref{2bhm}) as follows
\begin{eqnarray}
\label{H_spin}
    {\cal H} 
    &=& \sum_{j=1}^{L-1}\sum_{k=1}^{4}
    t_k
    \left(\sigma^+_{j,k} 
    \prod_{\ell=k+1}^4 \sigma_{j,\ell}^z
    \prod_{\ell=1}^{k-1} \sigma_{j+1,\ell}^z
    \sigma^-_{j+1,k}
    +{\rm H.c.}
    \right)\nonumber
        \\
    &+& \sum_{j=1}^{L}\left(
    U_{gg}\pi^0_{j,1} \pi^0_{j,2} + U_{ee} \pi^0_{j,3} \pi^0_{j,4}
    \right)-\sum_{j=1}^{L}\sum_{k=1}^{4}\mu_k  \pi^0_{j,k}\nonumber
    \\
        &+&
    V\sum_{j=1}^{L}
    \left(\pi^0_{j,1}\pi^0_{j,4}+\pi^0_{j,2}\pi^0_{j,3}\right)
    \nonumber
    \\
        \nonumber
    \\
    &+&
   (V-{V}_{\rm ex})
    \sum_{j=1}^{L}\left(
    \pi^0_{j,1}\pi^0_{j,3}+
    \pi^0_{j,2}\pi^0_{j,4}\right)
    \nonumber
    \\
        &-&
    V_{\rm ex}\sum_{j=1}^{L}\left(
    \sigma^+_{j,1}\sigma^-_{j,2}\sigma^-_{j,3}\sigma^+_{j,4}
    + {\rm H.c.}
    \right).
\end{eqnarray}

Being a sum of local terms of the range up to four consecutive spin sites, the Hamiltonian~\eqref{H_spin} can be written as a matrix product operator (MPO)~\cite{Pirvu2010a}, and treated with standard MPS numerical algorithms~\cite{McCulloch2007,itensor}.
The latter proceed by treating Eq.~(\ref{iterative}) \mc{with a fixed bond dimension $D$} as a variational ansatz  
\mc{ to minimize the energy}
$\langle\Psi|{\cal H}|\Psi\rangle / \langle\Psi|\Psi\rangle$.
\mc{The algorithm progresses by fixing all tensors but a single one $A_j$, and solving the resulting local problem~\cite{schollwock11}. The procedure is repeated one by one for all tensors, repeatedly sweeping over the chain until the energy converges. Because each local problem can be exactly solved, the energy decreases monotonically, and the algorithm is guaranteed to converge (even though it may do so to a local minimum), since the energy is lower-bounded. In practice, the sweeps are stopped when the relative change in the energy value is below a predetermined threshold. In our case, we fix this to be $10^{-8}$. 
Once the algorithm stops, we can improve the result by repeating the run with larger bond dimension, using the previous solution as starting point.
Comparing results with increasing bond dimension gives an estimate of their precision.
In all our simulations, we varied the bond dimension up to $260$. 
}

\mc{ The algorithm provides an explicit wave function of the form~\eqref{iterative} that approximates the ground-state. The expectation values of operators of interest can now be computed exactly in this state. 
}

As an additional verification of ED and MPS numerical results (as well as for a better understanding of physical mechanisms), at $n_g\approx1$ and $ n_e\approx 1$ we considered the strong coupling limit for the Hubbard model~\eqref{2bhm}, $t_{\gamma}\ll U_{\gamma\gamma'}$. In this limit, one can treat the tunneling as a perturbation and perform the Schrieffer-Wolff transformation to obtain an analytic form of the effective Hamiltonian. This aspect of studies will be discussed in more detail in Sec \ref{SC}.

\section{Results}

The band-structure calculations (similar to those performed in Ref.~\cite{Sotnikov2020PRR}) with the choice of parameters for the optical lattice specified in Sec.~\ref{sec:system} result in the values of the Hubbard pa\-ra\-me\-ters summarized in Table~\ref{tbl:Hparam}.
\begin{table}
\begin{tabular}{|c|cccccc|}
   \hline
               &{$t_{g}$($h\times$Hz)} & $t_{e}$  & $U_{gg}$ & $U_{ee}$ & $V$ & $V_{\rm ex}$ \\[1mm]
   \hline
   $^{173}$Yb & {160.1} 
   & 0.2591 & 9.238 & 18.13 & 37.031   & 25.646
   \\
   $^{171}$Yb & {161.9}
   & 0.2591 & 0 & 6.157 & 15.005   & -3.363
   \\
   $^{87}$Sr & {277.7}
   & 0.2591 & 4.16  & 9.727 & 5.724    & 2.439
   \\
   \hline
\end{tabular}
    \caption{Amplitudes of the Hubbard parameters for $^{173}$Yb, $^{171}$Yb, and $^{87}$Sr atoms in units of the tunneling amplitude $t_{g}$. }
    \label{tbl:Hparam}
\end{table}
Note that the inter-orbital interaction amplitudes $V$ and $V_{\rm ex}$ for $^{173}$Yb are additionally renormalized due to the fact that the ``bare'' amplitude $U_{eg}^+$ exceeds the band gap (see also Ref.~\cite{Sotnikov2020PRR} for details), while for other atoms all the amplitudes are moderate and obtained directly by means of Eq.~\eqref{Hubb_par}.

\begin{figure}
\begin{center}
    \includegraphics[width=\linewidth]{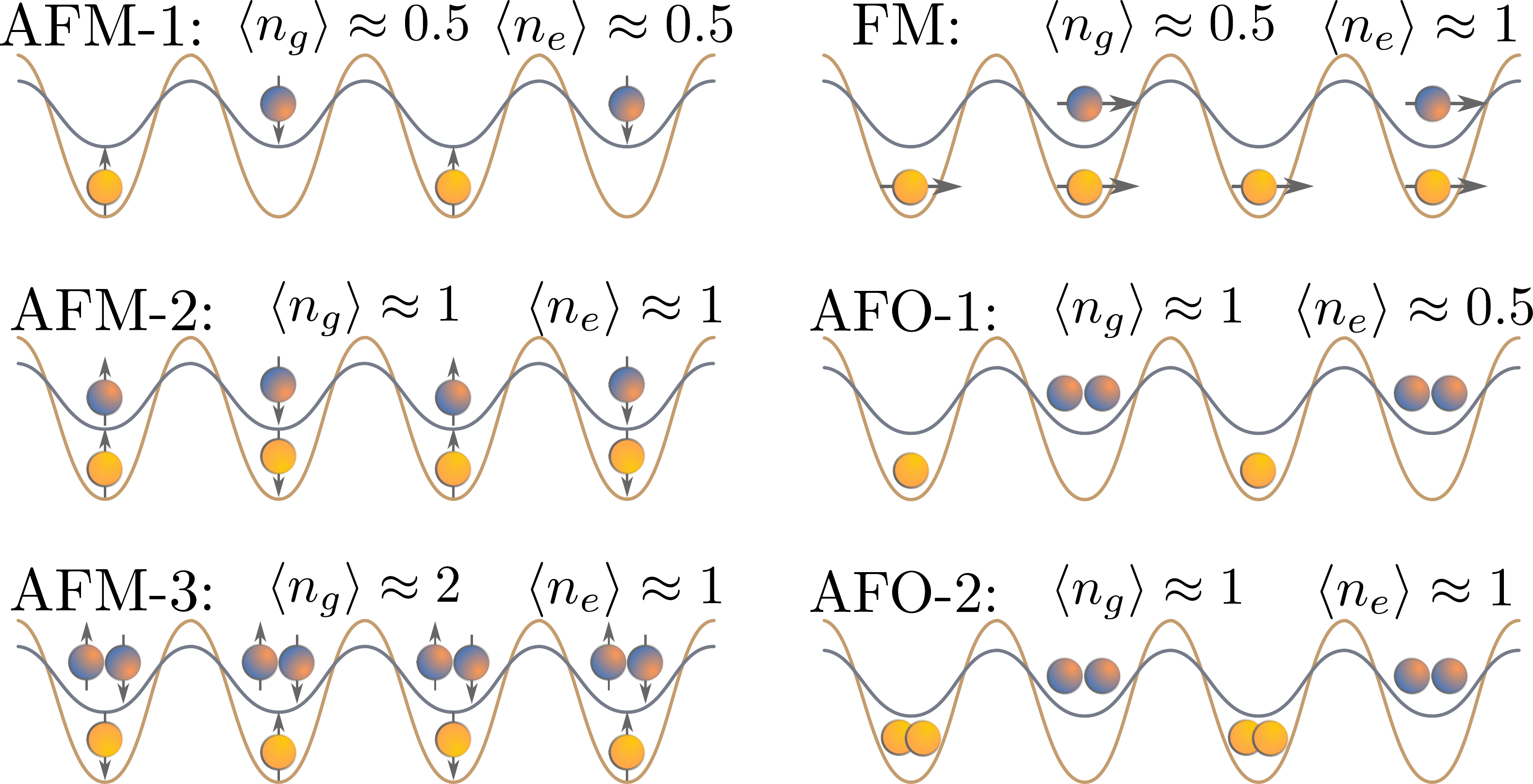}
    \caption{Schematic representation of magnetic and orbital ordering for particular average fillings of the lattice sites. Blue color corresponds to the ground-state atoms ($g$), while yellow color depicts the excited-state atoms ($e$).} 
    \label{regimes}
\end{center}
\end{figure}

In particular, for a gas of $^{173}$Yb atoms we observe a hierarchy of the interaction amplitudes similar to the one employed in recent theoretical studies with dynamical mean-field theory (DMFT) for a quasi-two-dimensional and three-dimensional geometries of SDL \cite{Sotnikov2020PRR, Sotnikov2020APPA}.
There, the authors pointed out a peculiar antiferrorbital (AFO) ordering instability in this system (also called as orbital density wave, see, e.g., Refs.~\cite{Nonne2010, Bois2015, Capponi2016}) among other strongly-correlated phases, antiferromagnetic (AFM) and ferromagnetic (FM), in particular (see also Fig.~\ref{regimes}).
Although DMFT is an approximate method, it is important to verify whether the main observations remain valid for a quasi-one-dimensional geometry of SDL with  the more accurate methodology employed here (see Sec.~\ref{sec:methods}).

\subsection{Spin-averaged local observables}
Due to the computational limitations mentioned in Sec.~\ref{sec:methods}, we perform ED calculations for a system size up to $L=5$.
Despite the limited size, the ED results already indicate se\-ve\-ral important features of the system under study.
Depending on the ratio between the densities of $g$ and $e$ atoms in the lattice, different correlations can be effectively enhanced in the density (or ``charge density'', if one uses an analogy to solid-state realizations), spin, and orbital sectors.
The ED approach also enables a straightforward temperature analysis and serves as an accurate control of the MPS approach. 

In order to find a trade-off
between calculation time and capturing all the relevant features of the system, while employing MPS we chose $L=20$ (and $D=260$) to represent the central and the most complete results of the study. 
\mc{We found that with these parameters, all the correlators discussed below are sufficiently converged, meaning that their sign does not change and they vary less than a few percent in relative value compared to the ones obtained with lower $D=220$, with most points (notice there are more than 800 points per panel) having actually much smaller errors $\lesssim 10^{-3}$.}
All the calculations were performed for the fixed number of $g$ and $e$ atoms
\footnote{This is realized by using the corresponding Lagrange multipliers while minimizing the ground-state energy of the Hamiltonian~\eqref{H_spin}}.
Note that below we focus mostly on spin-balanced configurations with the corresponding condition $N_\uparrow = N_\downarrow$ for the total number of particles $N_\sigma=\sum_{i\gamma}n_{i\gamma\sigma}$ in each spin state $\sigma$. It is worth mentioning that for the odd total number $N=N_{\uparrow}+N_\downarrow$  of atoms in the system, $N_{\uparrow}$ is set as rounding down of $N/2$ to the closest integer value.

We start our analysis with the on-site double occupancy $D_{gg}$ of $g$ atoms, which can be viewed as the global observable easily accessible in the experiments with ultracold multicomponent fermionic mixtures in the lattice. 
\as{
In particular, this can be detected for AEAs using a photoassociation resonance on the $^1$S$_0\to^3$P$_1$ intercombination
line. Two atoms can form a bound pair by absorbing a photon when the light is resonant with a bound state of the electronically excited molecule. Due to the short lifetime of the excited molecule, it will eventually decay and the released energy will cause a loss of the atom pair from the trap (see, e.g., Refs.~\cite{Tai2012Nat,Tusi2022} for more details).}
This observable is theoretically determined as $D_{gg}=\frac{1}{L}\sum\limits_{i}\langle n_{ig\uparrow}n_{ig\downarrow}\rangle$. 
As we will see below, $D_{gg}$ can be viewed as a good indicator of the onset of nearest-neighbor magnetic correlations in gases of $^{173}$Yb or $^{87}$Sr atoms, while for $^{171}$Yb there is no such correspondence.

\begin{figure}
\begin{center}
    \includegraphics[width=\linewidth]{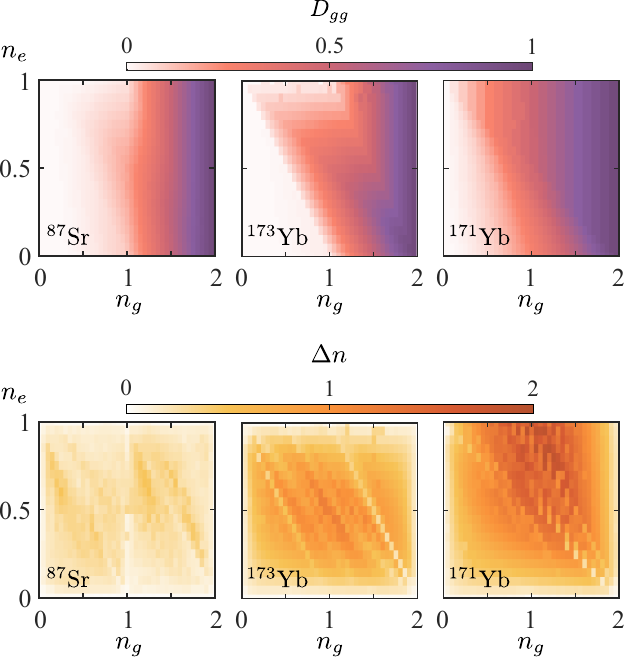}
    \caption{Site-averaged double occupancy of $g$ atoms $D_{gg}$ (the upper row) and density modulation $\Delta n$ (the lower row) depending on the filling of $g$ ($x$ axis) and $e$ ($y$ axis) orbital states, for three isotopes: $^{87}$Sr, $^{173}$Yb and $^{171}$Yb (from the left to the right side) and $L=20$ sites. The values of both observables are coded in colors.} 
    \label{double_charge}
\end{center}
\end{figure}

The dependence of the doubly-occupied sites with $g$ atoms on the variable densities $n_g$ and $n_e$ is shown in Fig.~\ref{double_charge} (upper row). 
Pauli exclusion principle imposes restrictions on the double occupancy $D_{gg}\leq1$ and the densities $n_{g,e}\leq2$.
Note that we further restrict the range of density of $e$ atoms, $n_e\leq1$, according to the experimental limitations connected with an increase of lossy collisions with a further growth of $n_e$ \cite{Gorshkov2010NP}.  
It is clearly visible that for the $^{173}$Yb isotope there is a strong suppression of the $D_{gg}$ at $n_g\approx n_e\approx1$.
The reason for this behavior lies in the hierarchy of the on-site in\-te\-rac\-tions. 
In comparison to $^{171}$Yb, where the intraorbital interaction amplitude for $g$ atoms vanishes ($U_{gg}\approx0$), for $^{173}$Yb  
the doubly-occupied sites would significantly increase the ground-state energy of the system.
The observed suppression of $D_{gg}$ close to $n_g\approx n_e\approx1$ is also related to the enhancement of the nearest-neighbor magnetic correlations, which are discussed in Sec~\ref{nn_corr} (see Fig.~\ref{SzSz}). Similar to $^{173}$Yb, in a gas of strontium-87 atoms one can observe qualitatively similar behavior of the double occupancy.   

In Fig.~\ref{double_charge} (lower row) we also analyze the density-wave modulation by calculating the site-averaged amplitude,
$
    \Delta n = \frac{1}{L}\sum_{i,\gamma}
    |\langle n_{i\gamma}\rangle - n_\gamma|.
$
\as{This quantity can also be measured in cold-atom systems by using an additional superlattice potential \cite{Schreiber2015}.
According to our theoretical analysis, it} demonstrates a different behavior to the double occupancy.
As we will see below, its enhancement can be used as an additional indicator of the orbital correlations ($^{171}$Yb and $^{173}$Yb), while its suppression can be attributed to the onset of antiferromagnetic  correlations in the Mott-insulating regimes with $n=1$ or $n=2$ ($^{87}$Sr and $^{173}$Yb).

\subsection{Spin and orbital resolved nearest-neighbor correlators} \label{nn_corr}

In this subsection, we discuss features of the nearest-neighbor correlators, i.e., the spin-spin 
$\langle {\bf S}_i \cdot{\bf S}_{i+1}\rangle$ and orbital-orbital $\langle {T}^z_i {T}^z_{i+1}\rangle$ ones. 
\as{These quantities can be experimenatally measured by means of the quantum gas microscope techniques developed for alkaline-earth(-like) atoms \cite{Miranda15,Yamamoto16, Young22}.}
The local spin operator contains contributions from both orbital flavors, ${\bf S}_i={\bf S}_{ig}+{\bf S}_{ie}$, where the orbital components ${\bf S}_{i\gamma}=(S_{i\gamma}^x,S_{i\gamma}^y,S_{i\gamma}^z)$ are expressed in terms of conventional spin-1/2 Pauli matrices as 
${S}^{r}_{i\gamma}=\frac{1}{2} c^\dagger_{i\gamma \tau } \sigma^r_{\tau\tau'} c^{}_{i\gamma \tau'}$ for $r=(x,y,z)$.
In turn, the orbital correlator is defined 
in terms of the operator 
$T^z_i= \frac{1}{2}\sum_{\tau=\uparrow,\downarrow} c^\dagger_{i\gamma \tau } \sigma^z_{\gamma\gamma'} c^{}_{i\gamma'\tau}$.

\begin{figure}
\begin{center}
    \includegraphics[width=\linewidth]{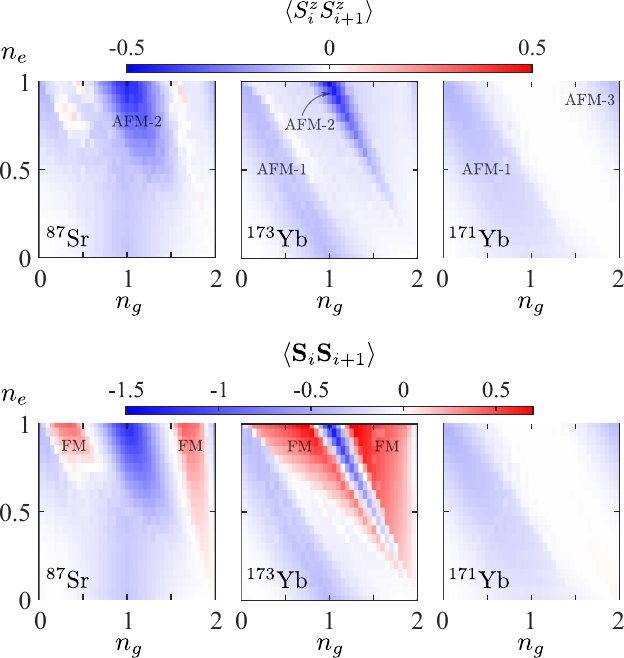}
    \caption{Dependencies of the site-averaged spin-spin correlators (coded in color), 
    \as{$\langle {S}^z_i {S}^z_{i+1}\rangle$} (upper row) and \as{$\langle {\bf S}_i \cdot{\bf S}_{i+1}\rangle$} (lower row),
    on the average fillings $n_g$ and $n_e$ for three isotopes: $^{87}$Sr, $^{173}$Yb and $^{171}$Yb (from left to right) obtained with the MPS approach at $L=20$.} 
    \label{SzSz}
\end{center}
\end{figure}
Figure~\ref{SzSz} presents the dependence of the site-averaged spin-spin correlators $\langle {\bf S}_i \cdot{\bf S}_{i+1}\rangle$ and $\langle {S}^z_i{S}^z_{i+1}\rangle$ on the densities of $g$ and $e$ atoms. One can observe that in case of $^{173}$Yb, the correlator $\langle {S}^z_i{S}^z_{i+1}\rangle$ reveals the antiferromagnetic ordering along diagonals $n_g+n_e=1$ and $n_g+n_e=2$ (AFM-1 and AFM-2 configurations, respectively; see also Fig.~\ref{regimes}), which is manifested by the negative value of $\langle {\bf S}_i \cdot{\bf S}_{i+1}\rangle$.
Note that particularly in these regions we observe a strong suppression of the double occupancy $D_{gg}$ (see Fig.~\ref{double_charge}).
In turn, the $^{87}$Sr system exhibits weaker AFM correlations along the same diagonals as $^{173}$Yb due to lower values of the interaction parameters, but with a similar correspondence in suppression of the $D_{gg}$ signal.
Surprisingly, a gas of $^{171}$Yb atoms with the AFM on-site Hund's coupling ($V_{\rm ex}<0$, see Table~\ref{tbl:Hparam}) does not demonstrate any AFM correlations at $n_g\approx n_e\approx1$. The reason for that
originates from the different hierarchy of the interaction amplitudes and thus a different ground state in the strong-coupling limit (see also Sec.~\ref{SC} for more details).

Next, one can notice that for both isotopes with ferromagnetic  Hund's coupling ($^{173}$Yb and $^{87}$Sr with $V_{\rm ex}>0$, see Table~\ref{tbl:Hparam}) there are certain regimes with a strong FM signal in the correlator $\langle {\bf S}_i \cdot{\bf S}_{i+1}\rangle$. 
This FM signal is almost absent in the correlator $\langle {S}^z_i{S}^z_{i+1}\rangle$ due 
to the constraint for finite size and zero total polarization, $N_\uparrow=N_\downarrow$.
In turn, due to the AFM exchange interaction ($V_{\rm ex}<0$) in the $^{171}$Yb system, no ferromagnetic correlations develop, which also results into direct correspondence between the depicted spin-spin correlators $\langle {\bf S}_i \cdot{\bf S}_{i+1}\rangle$ and $\langle {S}^z_i{S}^z_{i+1}\rangle$ in the whole diagram.

Therefore, the fillings $n_g$ and $n_e$, as well as the type of atomic isotope, determine four different magnetic orderings, depicted schematically in Fig. \ref{regimes}, that we label AFM-1 ($n_g+n_e\approx1$), AFM-2 ($n_g+n_e\approx2$), AFM-3 ($n_g+n_e\approx3$) and FM. 
We performed additional calculations in the regions $n_e>1$ (not shown in figures), which demonstrate that the spin-spin (as well as orbital-orbital) correlators are symmetric with respect to  reflections from the line $(n_g+n_e)=2$.
This fact is directly related to the particle-hole symmetry in both orbital flavors and can be useful for verification and control purposes.

\begin{figure}
\begin{center}
    \includegraphics[width=\linewidth]{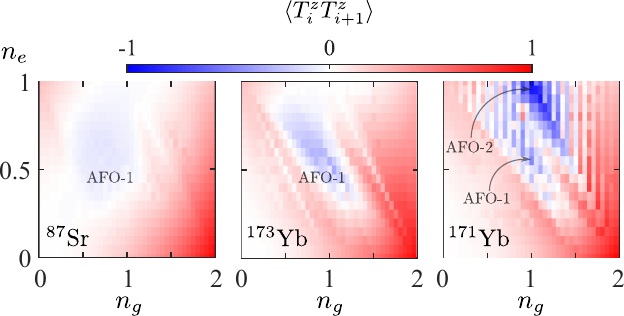}
    \caption{Dependencies of the site-averaged orbital-orbital correlators $\langle {T}^z_i {T}^z_{i+1}\rangle$ (coded in color) on the average fillings $n_g$ and $n_e$ for three isotopes: $^{87}$Sr, $^{173}$Yb and $^{171}$Yb (from left to right) obtained with the MPS approach at $L=20$.} 
    \label{TzTz}
\end{center}
\end{figure}
Finally, Fig.~\ref{TzTz} shows the dependence of the orbital-orbital correlators $\langle {T}^z_i {T}^z_{i+1}\rangle$ on the average densities of $g$ and $e$ atoms. In the case of $^{173}$Yb, one can observe antiferroorbital ordering around $n_g=1$ and $n_e=0.5$ (see also Fig.~\ref{regimes}), which is manifested by negative correlations.
Remarkably, the position and extent of this phase is in a good agreement with the previous DMFT studies of quasi-2D lattice systems \cite{Sotnikov2020PRR}.
The main feature of the AFO phase is the alternating occupation of neighboring lattice sites by atoms in different orbital states.
For illustrative purposes, the idealized configuration (that we name AFO-1) at $n_g=1$ and $n_e=0.5$ is shown in Fig.~\ref{regimes}.
Note that the AFO-like density modulations also emerge in the case of a gas of strontium atoms with the corresponding maximum of the signal at $n_g=1$ and $n_e=0.5$, but with a lower magnitude.
It is worth mentioning that the sharp transition features in the dependencies of $\langle {T}^z_i {T}^z_{i+1}\rangle$ for $^{173}$Yb and $^{171}$Yb at $n\approx1$, $n\approx1.5$, and $n\approx2$ in Fig.~\ref{TzTz}, as well as in the density-related local correlators (see Fig.~\ref{double_charge}), correspond to the transitions to the insulating regimes with the charge gap, which take place also in the thermodynamic limit.

Let us also discuss the dependence of the correlator $\langle {T}^z_i {T}^z_{i+1}\rangle$ for the case of $^{171}$Yb isotope shown in Fig.~\ref{TzTz}. 
In contrast to $^{173}$Yb and $^{87}$Sr atomic systems, one observes the strongest AFO signal at $n_g=n_e=1$  
(labeled as AFO-2 in Fig.~\ref{regimes}).
AFO-2 is a bipartite ordering similar to AFO-1. However, the main difference is that the neighboring lattice sites are occupied alternately by pairs of $g$ or $e$ atoms. 
The reason for the AFO instability (which completely suppresses the AFM correlations, see Fig.~\ref{SzSz}) in this particular regime for $^{171}$Yb system originates from the different hierarchy of the interaction amplitudes and thus a different ground state in the strong-coupling limit (see also Sec.~\ref{SC}). 
Note also the alternating vertical-stripe suppression features in the values of the correlator $\langle {T}^z_i {T}^z_{i+1}\rangle$ in Fig.~\ref{TzTz}. We ascribe these to the finite-size effects and vanishing $U_{gg}$. In particular, the suppression is observed at  odd values of the total number of $g$ atoms in the system $N_g$, 
when pairs of $g$ atoms cannot be any longer uniformly distributed along the chain (\as{e.g., }on  every second site).
With an increase of the system size these suppression features become less pronounced and we expect them to vanish in the thermodynamic limit ($L\to\infty$).
Let us also note that the AFO correlations are usually accompanied by sizeable density modulations (the charge-density wave) on the nearest-neighbor lattice sites. This can be concluded, in particular, from the corresponding comparison of Figs.~\ref{double_charge} and \ref{TzTz}.

\subsection{Strong-coupling limit at half filling}\label{SC}

Let us analyze in detail a regime with $n_g\approx1$ and $n_e\approx1$, when hopping processes can be viewed as a perturbation. The numerical ED and MPS results suggest strong correlations of different types in this region. 
In particular, the structure of these depends  on the atomic isotope: there are clear indications of the AFM correlations for $^{173}$Yb and $^{87}$Sr, while in the system consisting of $^{171}$Yb atoms the AFO correlations become the leading ones (see Figs.~\ref{SzSz} and \ref{TzTz}).
Since the interaction amplitudes are much larger than the hopping amplitudes for all three atomic species (see also Table~\ref{tbl:Hparam}), it is natural to employ the strong-coupling expansion.

To proceed, we restrict ourselves to two lattice sites and balanced spin configurations for both $g$ and $e$ components.
The atomic limit ($t_g=t_e=0$) already sets the different  lowest-energy states depending on the atomic isotope.
In particular, for the $^{171}$Yb atoms the ground state corresponds to the AFO-2 configuration shown in Fig.~\ref{regimes} with the zeroth-order contribution to the energy $E_0^{\rm AFO}=(U_{ee}+U_{gg})/2$ per lattice site.
At the same time, for other species, due to different hierarchy of the interaction amplitudes, the lowest-energy state is degenerate and formed by the local spin-triplet states ($S=1$, see Fig.~\ref{interactions}) consisting of pairs of $g$ and $e$ atoms on each lattice site with the zeroth-order contribution to the energy $E_0=(V+V_{\rm ex})$.  
This degeneracy is removed by accounting for the hopping processes and results in the AFM-2 configuration shown in Fig.~\ref{regimes}.

To verify the above statements and to estimate the characteristic  magnetic (or orbital) couplings, we apply  the Schrieffer-Wolff transformations \cite{Fazekas1999} and arrive at the following effective Hamiltonian at half filling:
\begin{eqnarray}\label{Heff_AFM}
    \mathcal{H}_{\rm eff}^{\rm AFM} &=&\mathlarger{\sum}_{\langle ij\rangle, \gamma\neq \gamma'} \frac{4 t_{\gamma}^2}{U_{\gamma\gamma}+V_{\rm ex}}  
    \Big({\bf S}_{i\gamma}\cdot{\bf S}_{j\gamma}-\frac{n_{i\gamma}n_{j\gamma}}{4}\Big) n_{i\gamma'}n_{j\gamma'}\nonumber 
    \\
    &&+ \mathcal{H}_{\rm int},
\end{eqnarray}
where the orbital-resolved spin-operators ${\bf S}_\gamma$ are defined as above (see Sec.~\ref{nn_corr}). 
Note that for the validity of this model it is necessary that $t_{\gamma}^2\ll({U_{\gamma\gamma}+V_{\rm ex}})$, which is guaranteed for the systems under study (see Table~\ref{tbl:Hparam}).

By performing a similar strong-coupling expansion for the AFO-2 configuration (see also Fig.~\ref{regimes}), we obtain the following effective model:
\begin{eqnarray}\label{Heff_AFO}
    \mathcal{H}_{\rm eff}^{\rm AFO}
    &=&\mathlarger{\sum}_{\langle ij\rangle}\mathlarger{\sum}_{\sigma\neq\sigma', \gamma\neq \gamma'} 
    \frac{4t_{\gamma}^2}{2V-U_{\gamma\gamma}-V_{\rm ex}}
    T_{i\sigma}^zT_{j\sigma}^z n_{i\gamma \sigma'}n_{j\gamma'\sigma'}\nonumber 
    \\
    &&+ \mathcal{H}_{\rm int}.
\end{eqnarray}
Here, the applicability of the model is related to the condition $t_{\gamma}^2\ll({2V-U_{\gamma\gamma}-V_{\rm ex}})$, which is also guaranteed for the systems under study (see Table~\ref{tbl:Hparam}).

We can conclude that both the AFO and the AFM correlations are mainly driven by the hopping of $g$ atoms (under assumption that $t_g>t_e$). 
At the same time, the denominators in the corresponding couplings are different due to the different structure of the ground and virtual states in different regimes.
Let us also note that we checked that the hierarchy of the ground-state energies at $n_g\approx1$ and $n_e\approx1$ remains unchanged for each atomic isotope with tuning of the polarizability ratio and the SDL depth.

\section{Conclusion}

We studied many-body correlations peculiar to the ground state of the gaseous systems consisting of interacting fermionic ytterbium or strontium atoms in state-dependent optical lattices.
Our theoretical analysis for a quasi-one dimensional geometry of SDL revealed a substantial number of distinct regimes with characteristic magnetic, orbital, and density correlations.
We calculated both single- and two-site (as well as the spin-averaged and spin-resolved) observables, which can be measured in the corresponding experimental realizations with ultracold atoms.
In particular, the obtained results are relevant not only for experiments with an access only to the averaged observables (e.g., double occupancy, density distribution, compressibility, etc), but also for experiments with the single-site resolution (quantum gas microscope) techniques in AEAs \cite{Miranda15,Yamamoto16, Young22}.

Although we restricted ourselves to certain values of the lattice depth and polarizability ratio, the comparison of different atomic isotopes provides useful information on how the necessary regimes can be approached and analyzed in different atomic systems.
Our results open also interesting directions toward realization of complex inhomogeneous systems, where the trap curvature can be adjusted to enhance one specific or several different phases in different spatial regions of the trap. 
Furthermore, the employed approaches can be extended to account for thermal effects and to perform the entropy analysis, which is valuable from the experimental point of view.
A good qualitative agreement of the results for $^{173}$Yb gas with Ref.~\cite{Sotnikov2020PRR} constitutes an indication that the main strongly-correlated regimes for all three atomic systems should remain stable and could be observed in the higher-dimensional systems at finite temperature.

\vspace{-2mm}
\begin{acknowledgments}
\vspace{-2mm}
The authors thank Nelson Darkwah Oppong and Ravindra Chhajlany for helpful discussions.
V.B., E.B., and A.S. acknowledge support from the National Research Foundation of Ukraine, Grant No.~0120U104963, the Ministry of Education and Science of Ukraine, Research Grant No.~0122U001575, and the National Academy of Sciences of Ukraine, Project No. 0121U108722.
M.C.B. was partially supported by the Deutsche Forschungsgemeinschaft (DFG, German Research Foundation) under Germany's Excellence Strategy -- EXC-2111 -- 390814868, and by the EU-QUANTERA project TNiSQ (BA 6059/1-1).
Access to computing and storage facilities provided by
the Poznan Supercomputing and Networking Center (EAGLE
cluster) is greatly appreciated.
\end{acknowledgments}

\paragraph*{Data availability.} The datasets generated and analysed during the current study are available in the arXiv repository,
\url{https://arxiv.org/src/2302.10854v2/anc}.

\bibliography{ael_correl}

\end{document}